%% file: main.tex
\documentclass[conference]{IEEEtran}
\usepackage[pdftex]{graphicx}
\graphicspath{{./}{./figures/}}
\DeclareGraphicsExtensions{.pdf,.jpeg,.png}

\usepackage{algorithm}
\usepackage{multirow}
\usepackage{verbatim}
\usepackage{amsmath}
\usepackage{listings}
\usepackage{xcolor}
\usepackage{color, colortbl}

\PassOptionsToPackage{hyphens}{url}
\usepackage[hyphens]{url}
\usepackage{hyperref}
\usepackage{booktabs}
\usepackage{clrscode}
\usepackage{color}
\usepackage{subfigure}
\definecolor{mygreen}{rgb}{0,0.6,0}
\newcommand{\specificthanks}[1]{\@fnsymbol{#1}}

\begin{document}

        
\title{Early Evaluation of Intel Optane Non-Volatile Memory with HPC I/O Workloads\thanks{notes on title}}


\newcommand{\email}[1]{\texttt{\small{#1}}}
\author{
  \begin{minipage}{5.0cm}
    \centering
    Kai Wu\\
    \email{kwu42@ucmerced.edu}
  \end{minipage}
 
 \begin{minipage}{5.0cm}
    \centering
     Frank Ober$^\dagger$\\
    \email{frank.ober@intel.com}
 \end{minipage}
 \\  \\
  \begin{minipage}{6cm}
    \centering
    Shari Hamlin$^\dagger$\\
    \email{shari.l.hamlin@intel.com}
  \end{minipage}
     \begin{minipage}{5.0cm}
    \centering
     Qiang Guan$^\star$\\
    \email{qguan@lanl.gov}
 \end{minipage}
 \begin{minipage}{5.0cm}
    \centering
    Dong Li\\
    \email{dli35@ucmerced.edu}
  \end{minipage} \\\\
University of California, Merced \qquad Intel Corp., USA$^\dagger$ \qquad Los Alamos National Lab$^\star$ 
}


\maketitle

\begin{abstract}
High performance computing (HPC) applications have a high requirement on storage speed and capacity. Non-volatile memory is a promising technology to replace traditional storage devices to improve HPC performance. Earlier in 2017, Intel and Micron released first NVM product -- Intel Optane SSDs. Optane is much faster and more durable than the traditional storage device. It creates a bridge to narrow the performance gap between DRAM and storage. But is the existing HPC I/O stack still suitable for new NVM devices like Intel Optane? How does HPC I/O workload perform with Intel Optane?  

In this paper, we analyze the performance of I/O intensive HPC applications with Optane as a block device and try to answer the above questions. We study the performance from three perspectives: (1) basic read and write bandwidth of Optane, (2) a performance comparison study between Optane and HDD, including checkpoint workload, MPI individual I/O vs. POSIX I/O, and MPI individual I/O vs. MPI collective I/O, and (3) the impact of Optane on the performance of a parallel file system, PVFS2. 

\end{abstract}

\renewcommand{\thefootnote}{\fnsymbol{footnote}}
\footnotetext[1]{LA-UR-17-27890}

\input text/Introduction   
\input text/Background_and_Motivation    
\input text/Evaluation_Setup 
\input text/Evaluation_results  
\input text/Related_work.tex
\input text/Conclusion_and_future_work   


\bibliographystyle{IEEEtran}
\bibliography{li,kai}
%

\end{document}

%% file: text/Introduction.tex
\section{Introduction}
The modern HPC applications can be very demanding for storage capacity and performance. For example, Blue Brain project aims to simulate the human brain with a daunting 100PB memory that needs to be revisited by the solver at every time step; The cosmology simulation studying Q continuum works on 2PM per simulation. 
To meet storage and performance requirement of those HPC applications, 
non-violate memory (NVM) technology is a promising building block for future HPC.

According to some recent literatures~\cite{NVMDB}, the emerging NVM techniques, such as phase change memory (PCM)~\cite{pcm:micro2010} and STT-RAM~\cite{str-ram:VLSI2008}, have shown many interesting performance characteristics. First, the performance of NVM is much better than that of a hard drive, and even closer to that of DRAM. Second, NVM has better density than DRAM while remaining non-volatile. Third, NVM has byte addressability which can largely improve load/store efficiency. Table~\ref{tab:nvm_perf} summarizes the features of different NVM techniques and compares them to the traditional DRAM and storage technologies.

\begin{table}[h]
        \begin{center}
            \caption{Memory Technology Summary~\cite{NVMDB}}
      \label{tab:nvm_perf}
    \begin{tabular}{|p{0.8cm}|p{1.2cm}|p{1.2cm}|p{1.5cm}|p{1.5cm}|}
     \hline
          & \textbf{Read time (ns)} & \textbf{Write time (ns)} &  \textbf{Read BW (MB/s)} & \textbf{Write BW (MB/s)}   \\ \hline
     DRAM       & 10    & 10 & 1,000 & 900        \\\hline
     PCRAM        & 20-200 & 80-$10^{4}$ & 200-800 & 100-800       \\ \hline
     SLC Flash       & $10^{4}$-$10^{5}$   & $10^{4}$-$10^{7}$ & 0.1 & $10^{-3}$-$10^{-1}$            \\ \hline
     ReRAM			 &  5-$10^{5}$ & 5-$10^{8}$ & 1-1000 & 0.1-1000     \\  \hline
     Hard drive      & $10^{6}$   & $10^{6}$   & 50-120 & 50-120   \\ \hline
     \end{tabular}
      \end{center}
\end{table}

Earlier in 2017, Intel and Micron released their joint NVM product Intel Optane SSD. 
Optane is based on 3D Xpoint technology. The 3D XPoint is 1000x faster and 1000x more endurable than NAND flash memory. 
To leverage Optane for future HPC, we must understand the performance impact of Optane on HPC applications. In this paper, we study the Optane performance with HPC applications under various test environments. This is the first work to evaluate performance of Intel Optane in the HPC field.   
%
%
%
In this paper, we conduct the following performance study. We first test Optane with Intel Open Storage Toolkit. We get preliminary understanding on this new NVM device's performance (particular read/write latency and bandwidth), and compare it with the performance of a traditional hard drive. Then, we analyze the performance of some I/O intensive HPC applications with Optane as a high-speed block device. We particularly compare the performance of MPI I/O and POSIX I/O based on Opetane. 
Furthermore, we compare the performance of two different MPI I/O techniques, MPI individual I/O and collective I/O. Collective I/O is an optimization technique based on the assumption of poor I/O performance. We want to know that whether it is still necessary to use MPI collective I/O for Optane.
Lastly, we investigate MPI I/O performance on Optane with a a parallel file system, PVFS. PVFS is widely used in HPC. How does Optane device affect the performance of a parallel file system? We study the performance with different configurations of PVFS. 

The rest of the paper is organized as follows. In Section 2 we present background information for this paper. Section 3 describes our test environment and benchmarks we use for evaluation. A detailed evaluation of Intel Optane on HPC I/O workloads is shown in Section 4. Finally, in Section 5 we conclude this paper and discuss our future work.

%% file: text/Background_and_Motivation.tex
\section{Background}
\subsection{Non-violate memory usage model}
\begin{figure}
\centering
\includegraphics[width=0.48\textwidth, height=0.24\textheight]{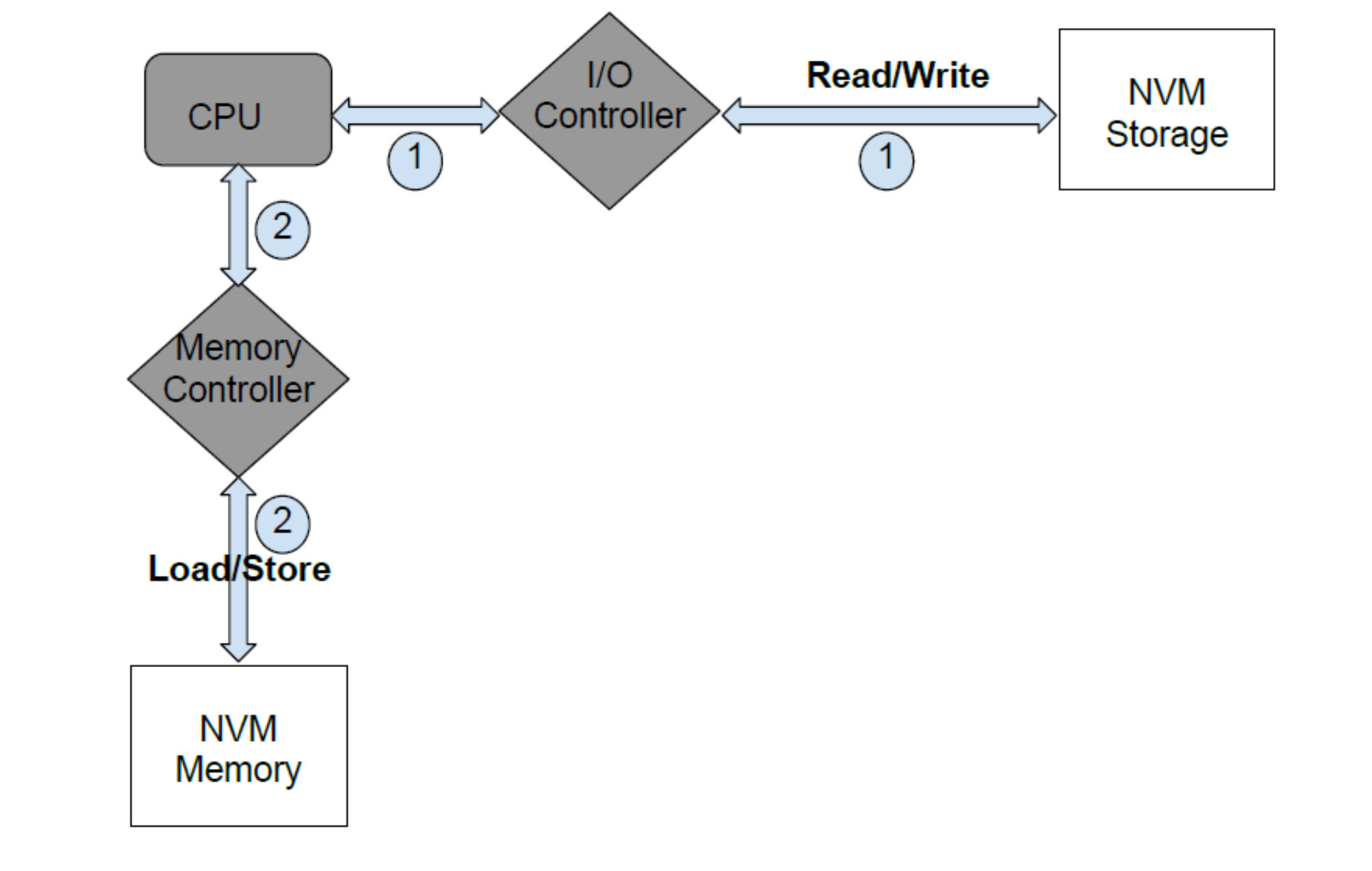}
\caption{Two types of NVM usage model}
\label{fig:hie}
\end{figure}

There are two major approaches to integrate NVM into the system. One approach is based on a memory-based model, and the other is based on a storage-based model. Figure~\ref{fig:hie} depicts the two approaches. 

(1) {\bf Storage-based Model.} Similar to traditional HDD and SSD, NVM is used as a block device in this usage model. The data path 1 in Figure~\ref{fig:hie} depicts such usage model. 
All operations including read and write are managed by I/O controller. Users can use regular block I/O interface, such as PCI-e or SATA, to access such storage-based NVM. However, using the storage-based model cannot easily achieve byte-addressability offered by NVM. 
Optane offers us an opportunity to use NVM based on the storage-based model.

(2) {\bf Memory-based Model.} In this model, NVM is attached to a high-speed memory bus as DIMMs. All load and store instructions are operated by the memory controller. The data path 2 in Figure~\ref{fig:hie} depicts such usage model. 

\subsection{Intel Optane Memory}
Optane Memory~\cite{Optane:intelweb, Optane:blogreview} is a 3D XPoint non-volatile memory jointly developed by Intel and Micron. The 3D XPoint technology is a new class of non-volatile memory that 
provides unprecedented performance. It can provide up to 1,000 times lower latency at the media level and many times greater endurance than NAND. 
The 3D XPoint technology can deliver game-changing performance for data-intensive applications. Its ability to enable high-speed, high-capacity data storage creates new possibilities for system architects to enable entirely new applications.

%% file: text/Evaluation_Setup.tex
\section{EXPERIMENTAL METHODOLOGY}
\subsection{Benchmarks}
In our evaluation tests, we use four benchmarks, which are Intel Open Storage Toolkit, HACC-IO, IOR, and mpi-io-test. The first three benchmarks are tested under a network file system (NFS) environment, and the fourth one is tested in a Parallel Virtual File System (PVFS) environment.

(1) {\bf Intel® Open Storage Toolkit.} This toolkit is used to measure I/O bandwidth, latency and IOPS of block device for different I/O workloads. The Intel Open Storage Toolkit supports various user configurable options, such as read/write ratio, random/sequential ratio, and request size.

(2) {\bf HACC-IO Benchmark.} This benchmark is the I/O kernel of HACC (an HPC application based on N-body simulation). It has random I/O write operations with all-to-all communication patterns. 

(3) {\bf IOR Benchmark.} IOR is a popular benchmark to study parallel I/O. It leverages the scalability of MPI to easily and accurately calculate the aggregate bandwidth of a user-defined number of client machines. IOR supports POSIX, MPI-IO (including individual I/O and collective I/O), and HDF5 I/O interfaces. Using IOR, the user can easily configure different I/O patterns, including ``sequential'' and ``random offset'' file access patterns. We use the sequential pattern in our tests.

(4) {\bf MPI-IO-test Benchmark.} This benchmark is shipped with PVFS2. It is used to test MPI I/O and functionality implemented by ROMIO~\cite{romio:mpich}. mpi-io-test can produce aggregate bandwidth numbers by varying block sizes, number of processors, number of iterations and MPI I/O methods (independent I/O or collective I/O). 

\subsection{PVFS}
PVFS can provide high performance access in a large-scale cluster computing environment. It is modularity, flexibility and with ROMIO support. In a PVFS cluster architecture, each node is used as one or more of: compute node, I/O node and  metadata node. I/O nodes hold file data; metadata nodes hold metadata include stat-info, attributes, and datafile-handles as well as directory-entries; compute nodes run applications that utilize the file system by sending requests to the servers over the network. 
We use PVFS2~\cite{pvfs2:linuxworld2004}, the second version of PVFS, for our tests.

\subsection{Evaluation Environment}
We have two experimental environment setups: one is with NFS, and the other is with PVFS2. The NFS environment has four nodes and each node has an Intel E52690v4  processor, 128 GB memory, 1 TB regular hard drive (attached by SATA), and 370 GB Optane SSD (attached by M.2 PCIe).  The PVFS environment has six nodes which have the same hardware configuration as the four nodes in the NFS environment. All nodes are provided by Intel, and we only compare the performance of Optane device and a regular hard drive (noted as ''HDD`` in this paper). We cannot compare the performance of Optane and SSD, because there is no SSD device on those nodes. All of the nodes use Linux kernel 3.10. We use MPICH-2.6 for MPI. We use 10GB Ethernet in our test environment.

%% file: text/Evaluation_results.tex
\section{PERFORMANCE STUDY}

\subsection{Basic Performance}
\label{sec:basic_perf}
We measure the raw bandwidth of Optann and HDD with Intel Open Storage Toolkit on a single node. Table~\ref{tab:basic_perf} shows the results. 
The table reveals that both read and write bandwidths of Optane device are much higher than those of HDD. For example, Optane achieves 357\% and 847\% of HDD bandwidth for random write and random read, respectively. 
We also notice that Optane has no big performance difference between 
random and sequential read/write. This is quite different from HDD that has big performance difference between random and sequential read/write.

\begin{table}
        \begin{center}
            \caption{Read and Write Performance (MB/s) on raw block device}
      \label{tab:basic_perf}
    \begin{tabular}{|p{0.8cm}|p{1.2cm}|p{1.2cm}|p{1.5cm}|p{1.5cm}|}
     \hline
 & \textbf{Write Random Bandwidth} & \textbf{Write Sequential Bandwidth} &  \textbf{Read Random Bandwidth} & 
 \textbf{Read Sequential Bandwidth}
 \\ \hline
     Optane     & 2174.08 & 2172.62 & 2286.15 & 2568.53 
 \\\hline
     HDD        & 6.08 & 200.25 & 2.7 & 204.30      
     \\ \hline
     \end{tabular}
      \end{center}
\end{table}


\subsection{Checkpoint Workload Performance}
\begin{figure}[!t]
    \centering
    \includegraphics[width=0.48\textwidth, height=0.2\textheight]{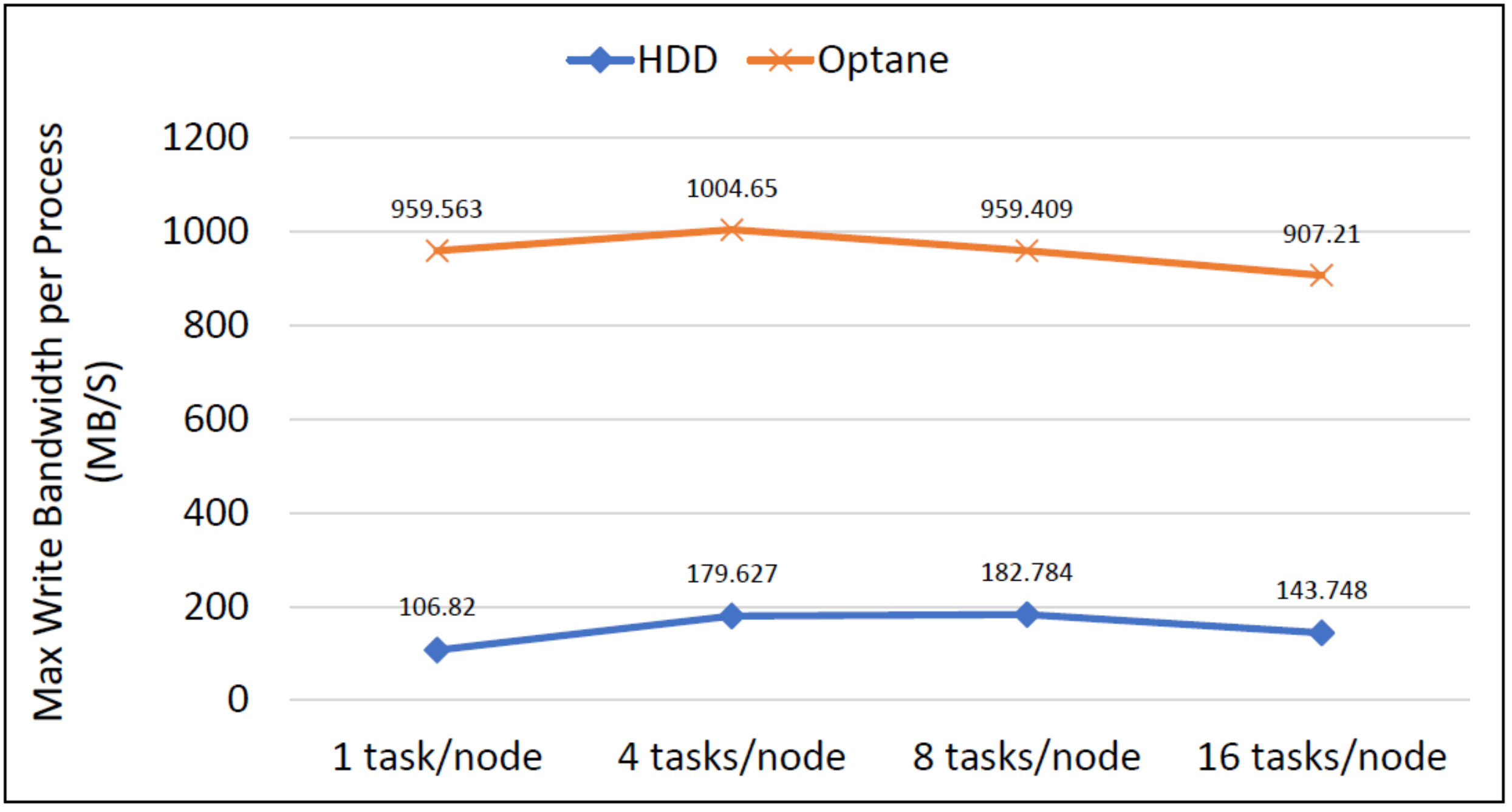}
    \caption{Performance evaluation with HACC-IO Benchmark}
    \label{fig:hacc-io}
\end{figure}

Checkpoint is a common fault tolerance mechanism in current production supercomputers and industrial data centers. In order to evaluate checkpoint performance on Optane, we use HACC-IO benchmark. 
In our tests, we use three nodes as compute node and one node as storage node. The compute nodes access storage nodes through NFS. 
We change the number of MPI tasks per compute node in our tests. Figure~\ref{fig:hacc-io} shows the results. 

We first notice that running HACC, the performance of Optane device is still much better than that of HDD, but the performance difference between Optane and HDD is smaller than the bandwidth difference we get in the basic performance evaluation (Section~\ref{sec:basic_perf}). 
The maximum performance difference between Optane and HDD when running HACC is 560\%, while the bandwidth difference between Optane and HDD in Section~\ref{sec:basic_perf} is at least 1000\%. 
We attribute such smaller performance difference to the network effect when doing remote I/O with NFS. The network latency is typically longer than I/O access latency in our test platforms. Hence the network becomes a performance bottleneck, overshadowing the performance benefit provided by Optane.

Another interesting result from Figure~\ref{fig:hacc-io} is that the Optane performance seems not very scalable in our tests. Ideally, the maximum bandwidth per MPI task shown in Figure~\ref{fig:hacc-io} should remain stable as we increase the number of MPI tasks per node. However, we see slightly decrease of the bandwidth per MPI task when the number of MPI tasks is larger than four. 
%
%

\subsection{POSIX I/O and MPI Individual I/O}
POSIX I/O is widely used in HPC, and MPI I/O adds another layer on top of POSIX I/O to improve the performance of parallel I/O. 
The functionality of MPI I/O includes data validation and data re-arrangement to achieve better I/O performance. However, MPI I/O introduces overhead. In the traditional storage system, such overhead is much smaller than the I/O performance benefit, justifying the necessity of using MPI I/O. However, given the excellent performance of Optane, the I/O performance benefit of MPI I/O could become smaller.
We expect that the overhead of MPI I/O will become more pronounced. 

\begin{figure}
\centering
\includegraphics[width=0.48\textwidth, height=0.2\textheight]{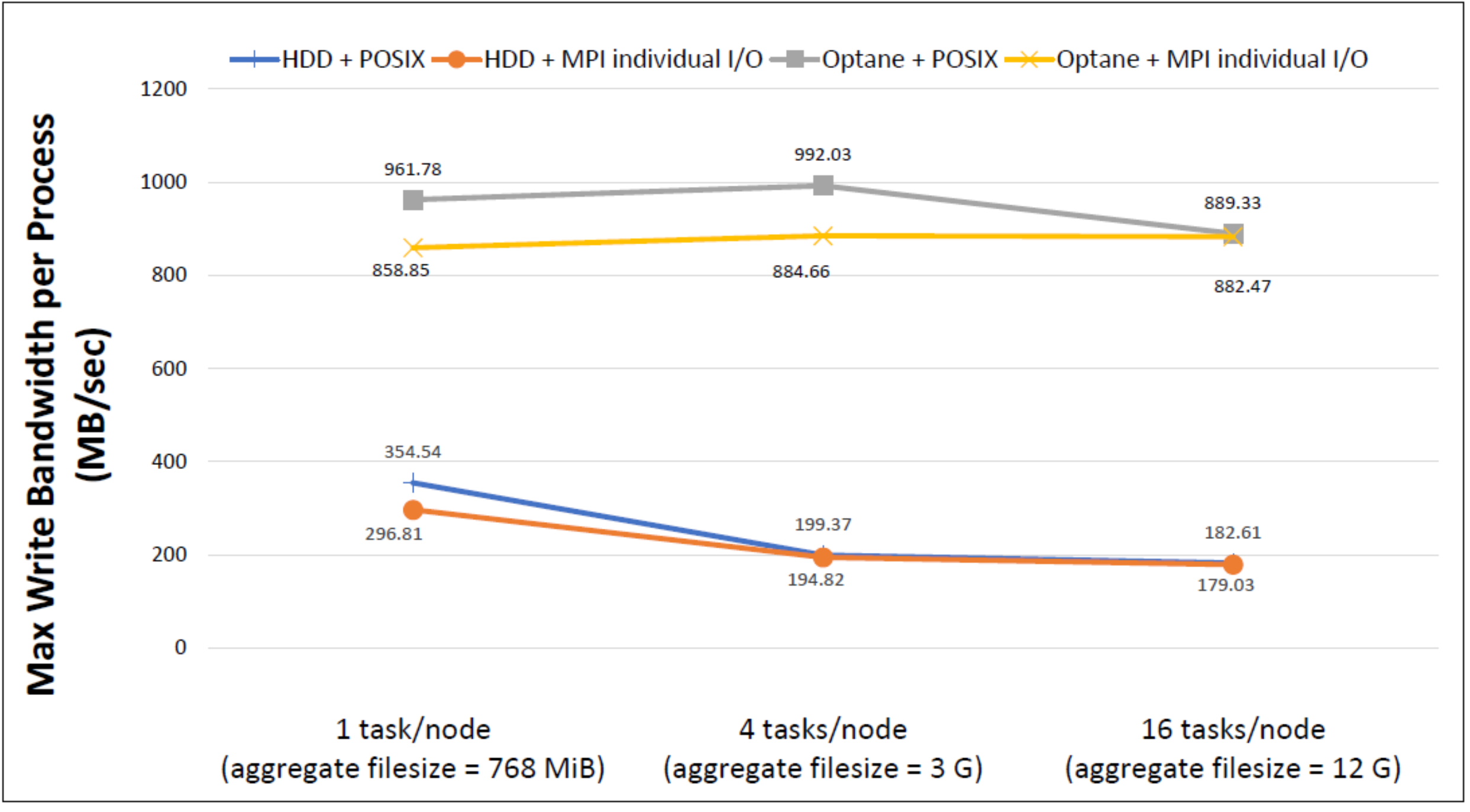}
\caption{Performance evaluation with IOR benchmark (POSIX I/O vs. MPI individual I/O)}
\label{fig:posix_mpi}
\end{figure}

In this test, we run IOR benchmark with three compute nodes and one storage node, 
and evaluate the performance of POSIX I/O and MPI individual I/O. 
We use NFS for remote storage access. 
The results are shown in Figure~\ref{fig:posix_mpi}. 
We find that when the number of MPI tasks per node is smaller (i.e., 1 or 4), the performance difference between POSIX I/O and MPI individual I/O on Optane is larger than that on HDD. This result is consistent with our expectation: the overhead of MPI I/O is more pronounced on Optane. 
However, when the number of MPI tasks per node is larger (i.e., 16), we see almost no performance difference between POSIX I/O and MPI individual I/O.
We attribute such smaller performance difference to the increasing performance benefit of MPI individual I/O when the number of MPI tasks per node is larger. 

\subsection{MPI Collective I/O and MPI Individual I/O}
\label{sec:collective_io_individual_io}
We compare performance of two MPI I/O methods, MPI collective I/O and MPI individual I/O, with IOR benchmark. 
We run IOR benchmark with three compute nodes and one storage node, 
and use NFS for remote storage access. 
MPI collective I/O is an optimization of MPI individual I/O.
MPI collective I/O re-arranges data distribution between MPI tasks before I/O operations to improve I/O performance. MPI collective I/O has two phases, i.e., data shuffle and I/O operation. 
By re-arranging data in the data shuffle phase, MPI collective I/O can combine multiple I/O operations, reduce the number of I/O transactions, and avoid unnecessary data fetching. However, the data shuffle brings overhead when re-arranging data between MPI tasks. In the traditional storage system, such overhead can be easily overweighted by the performance benefit because of the bad performance of the storage system. However, with Optane,  we expect that the data shuffle overhead cannot be easily overshadowed by the performance benefit. Hence, MPI collective I/O may become an unnecessary performance optimization. 

\begin{figure}
    \centering
    \includegraphics[width=0.48\textwidth, height=0.2\textheight]{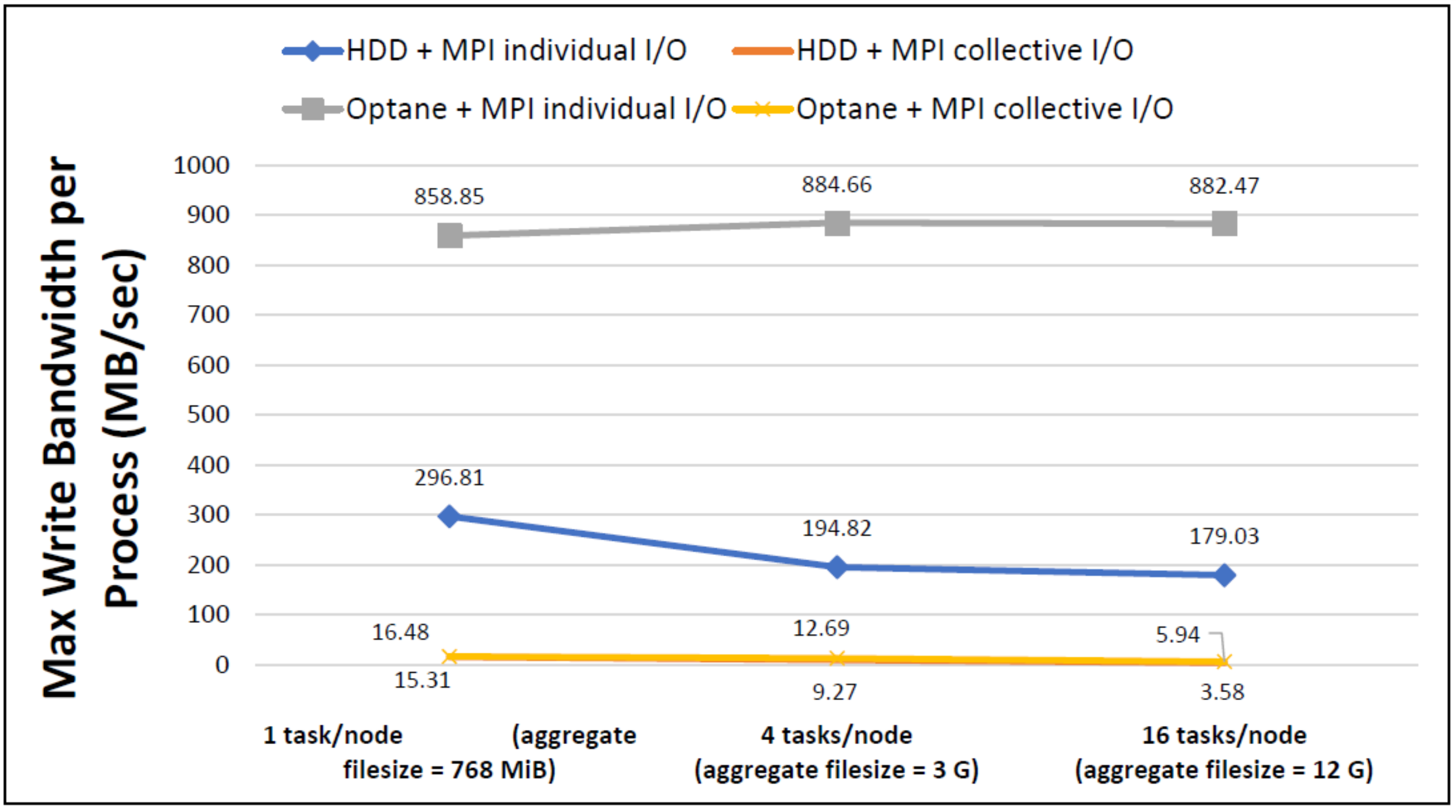}
    \caption{Performance evaluation with IOR benchmark (MPI individual I/O vs. MPI collective I/O)}
    \label{fig:IOR individual I/O collective I/O}
\end{figure}

Figure~\ref{fig:IOR individual I/O collective I/O} shows the results. Surprisingly, we do not see any performance benefit of MPI collective I/O in any test (neither Optane nor HDD), and MPI collective I/O always performs worse than MPI individual I/O. We attribute such result to the bad network performance in our test platform. The network performance becomes a bottleneck in the data shuffle phase. No matter how much I/O performance benefit MPI collective I/O can bring, it is always smaller than the overhead of the data shuffle. 

\subsection{Performance Evaluation with PVFS}
In this section, we change the configurations of PVFS, and then evaluate the performance of Optane. To use PVFS, we need to have I/O node, metadata node, and 
compute node. We change the numbers of those nodes in our evaluation. 
We use mpi-io-test during the evaluation. 

\begin{figure}[!t]
    \centering
    \includegraphics[width=0.48\textwidth, height=0.2\textheight]{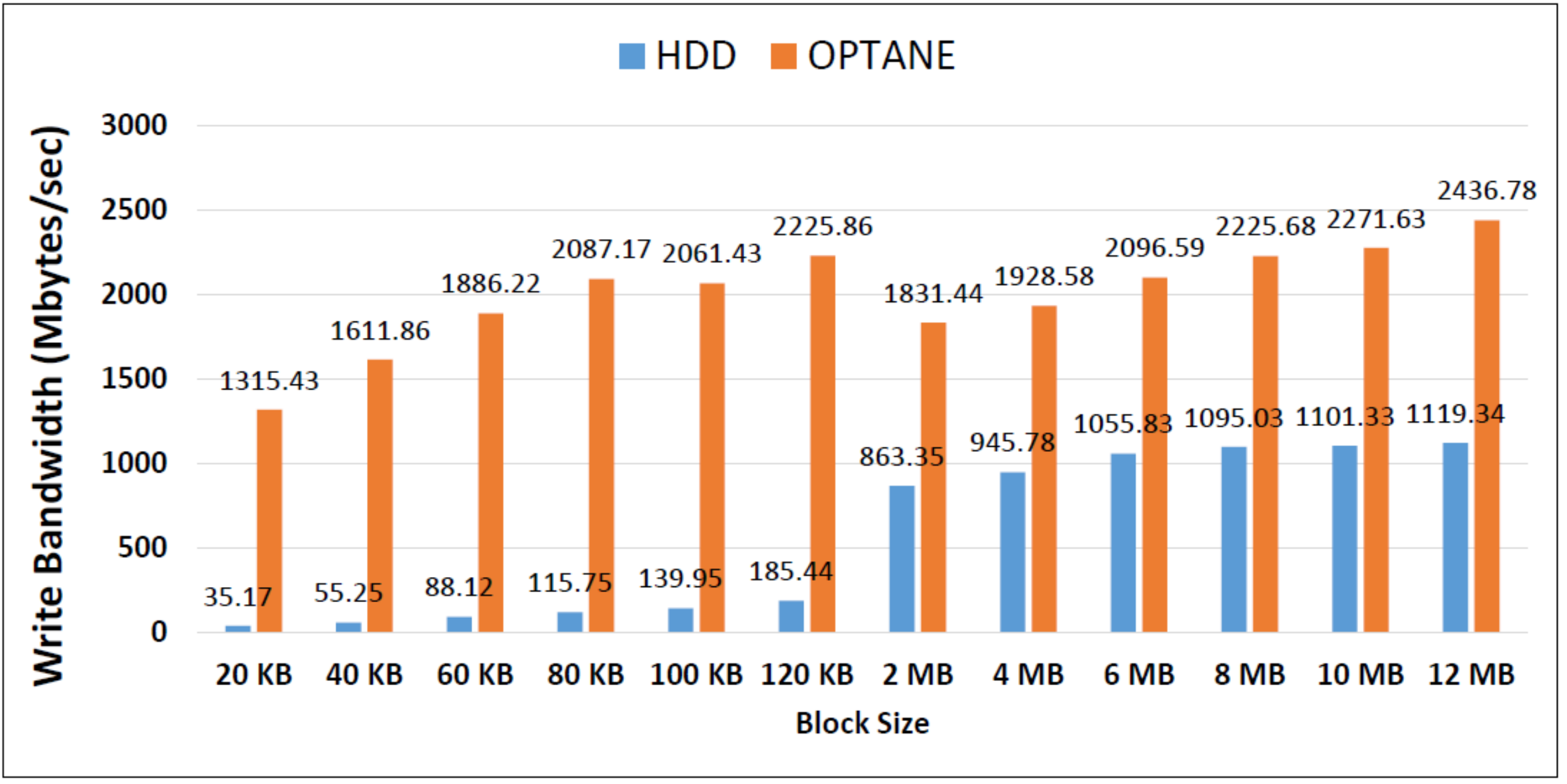}
    \caption{PVFS2 performance sensitivity to data block size on Optane}
    \label{fig:sensitivity_study_block_size}
\end{figure}

We first change the data block size and study the performance sensitivity of Optane to the data block size. We use one compute node, three I/O nodes and one metadata node for this test. We use 24 clients on the compute node. Figure~\ref{fig:sensitivity_study_block_size} show the results. 
The results show that write bandwidth of HDD continues growing when the block size increases. The similar story happens on Optane, but there is an interesting performance drop as we increase the block size from 120 KB to 2MB. We suspect such drop is because of the PVFS data alignment problem across multiple I/O nodes, but this performance drop needs to be further investigated. 


\begin{figure}
    \centering
    \includegraphics[width=0.48\textwidth, height=0.2\textheight]{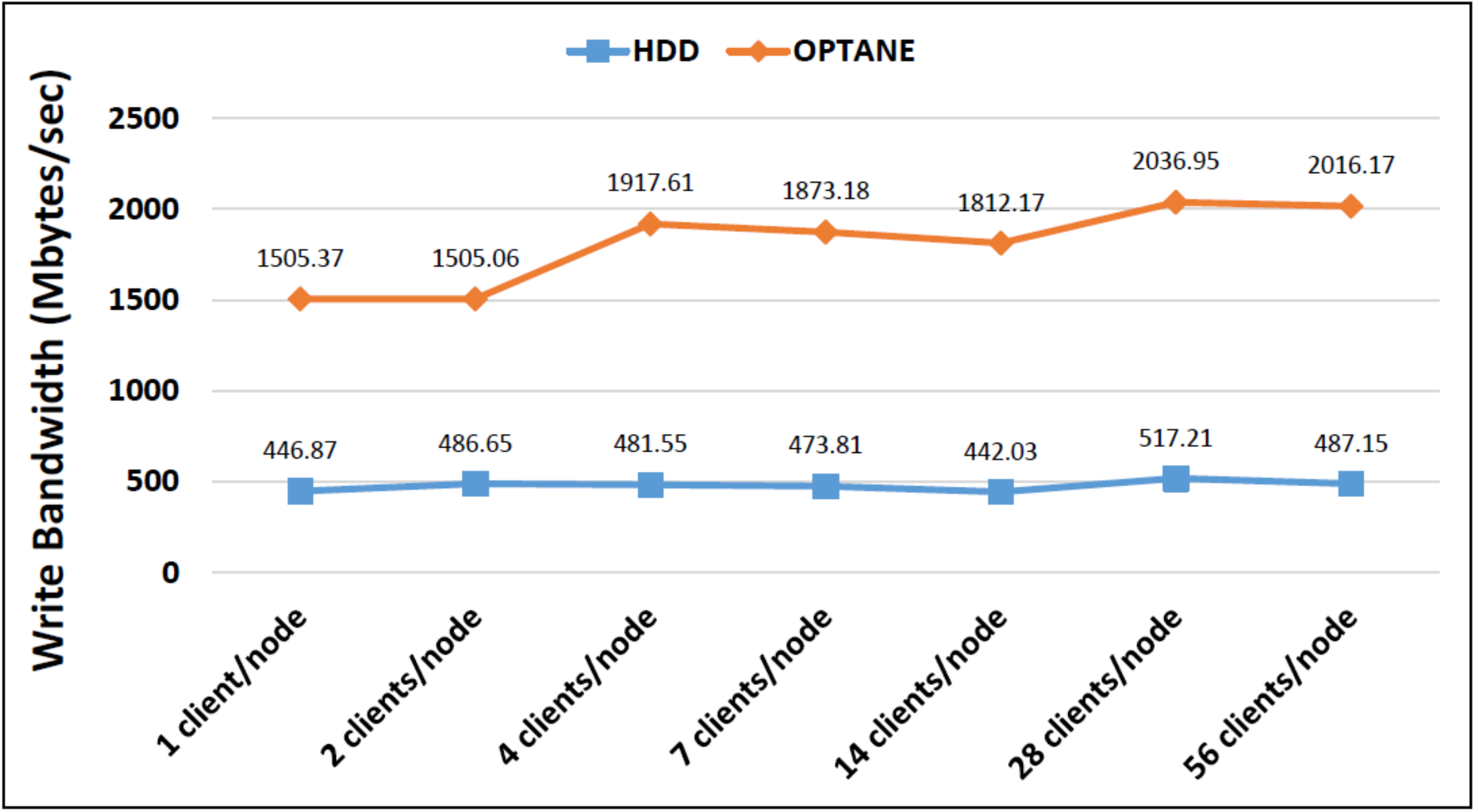}
    \caption{PVFS2 performance sensitivity to the number of clients per compute node}
    \label{fig:sensitivity_to_num_clients}
\end{figure}

We further study the performance sensitivity of Optane to the number of clients per compute node. We use two compute nodes, three I/O nodes and one metadata node for this test. Figure~\ref{fig:sensitivity_to_num_clients} show the results.
We notice that Optane has pretty good scalability. In particular, as we increase the number of clients, the write bandwidth keeps increasing. However, HDD does not have such good scalability: the write bandwidth remains relatively stable. 

\begin{figure}[!t]
    \centering
    \includegraphics[width=0.48\textwidth, height=0.2\textheight]{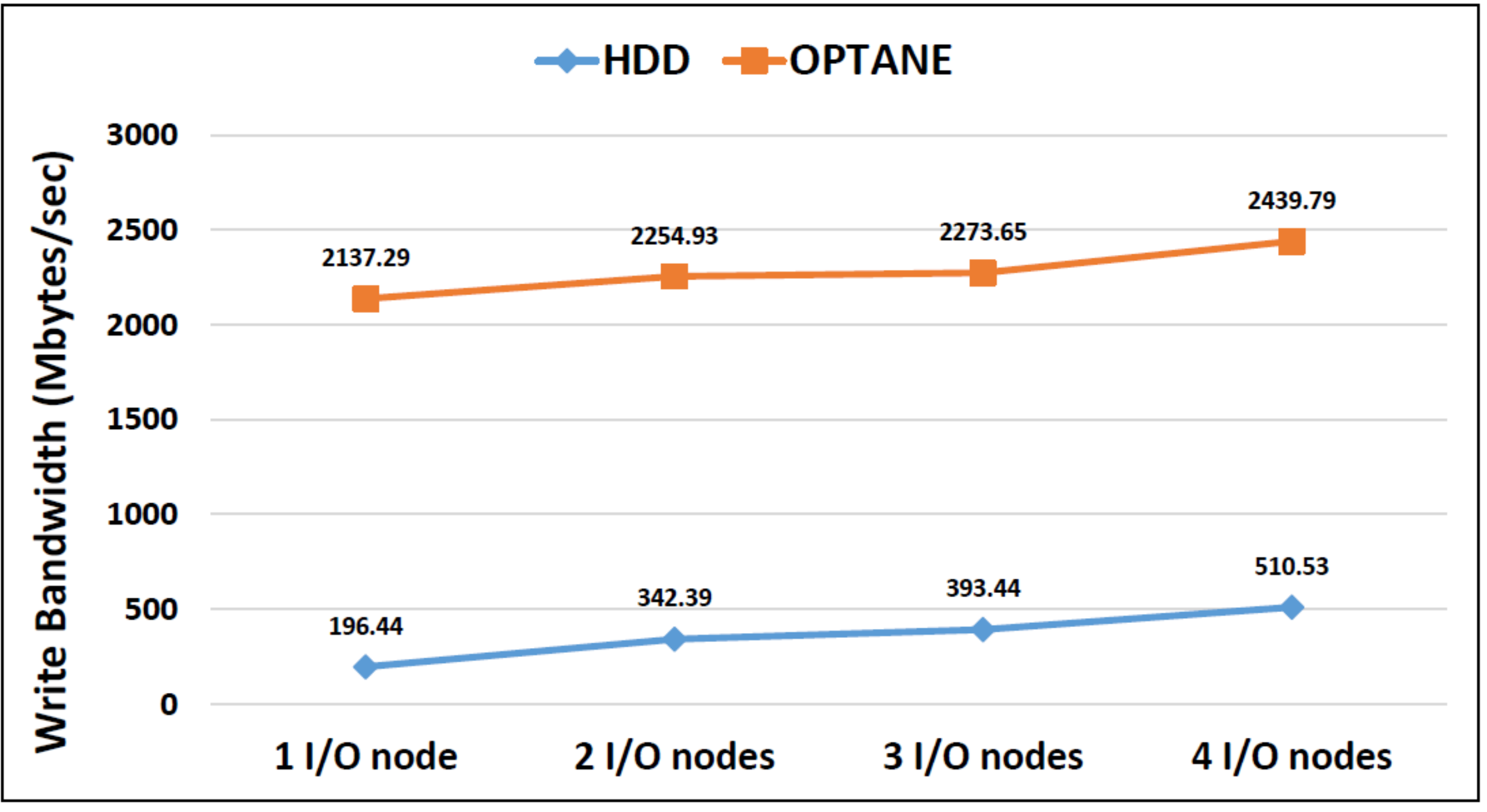}
    \caption{PVFS2 performance sensitivity to the number of of I/O nodes}
    \label{fig:sensitivity_study_to_number_io_node}
\end{figure}

We then study the performance sensitivity of Optane to the number of I/O nodes. We use one compute node and one metadata node for this test. We use 24 clients on the compute node. Figure~\ref{fig:sensitivity_study_to_number_io_node} show the results.
We notice that write bandwidth of both Optane and HDD increases as we increase the number of I/O nodes. However, their bandwidths increase at different rates.
For Optane, the bandwidth increases less than 15\%, while for HDD,
the bandwidth increases 2.6x. 
This indicates that for Optane with better I/O performance, the storage bandwidth
may not be the major performance bottleneck.
This performance result for Optane is different from HDD which suffers from the limited storage bandwidth. 


\begin{figure}[!t]
    \centering
    \includegraphics[width=0.48\textwidth, height=0.2\textheight]{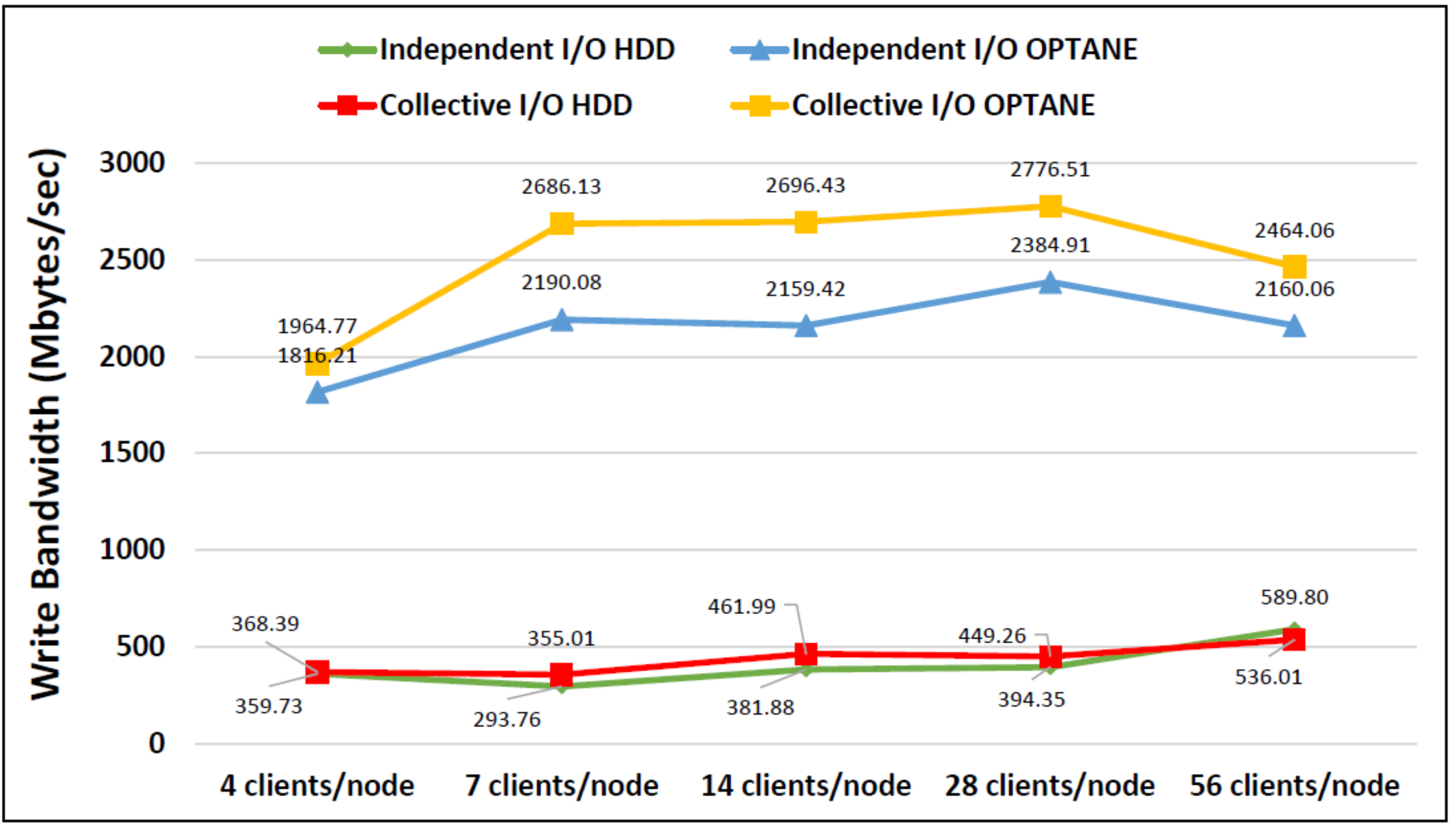}
    \caption{Comparing the performance of MPI individual I/O and MPI collective I/O with PVFS2}
    \label{fig:collective_io_individual_io}
\end{figure}

Lastly, we compare the performance of MPI individual I/O and MPI collective I/O on PVFS2. We use three I/O nodes, one metadata node and one compute node (24 clients on the compute node) for this tests. In Section~\ref{sec:collective_io_individual_io}, we have compared the performance of MPI individual I/O and MPI collective I/O in an NFS-based test environment, and find that MPI individual I/O always performs better than MPI collective I/O because of the bad network performance.
However, in the PVFS2-based test environment, we find a different story. 
MPI Collective I/O performs better than MPI Independent I/O on both Optane and HDD (see Figure~\ref{fig:collective_io_individual_io}). 
We reason that PVFS2 uses some performance optimization techniques to reduce the negative impact of network communication. The reduction of the impact of network communication reduces the overhead of the data shuffle which makes the benefit of MPI collective I/O more pronounced. 

%% file: text/Related_work.tex
\section{Related Work}
\label{sec:related_work}

\subsection{Non-Volatile Memory}
Non-volatile memory (NVM) technology is under quick development and has attracted a large body of research. A comprehensive survey about NVM can be found in a prior study~\cite{NVMDB}. There are some earlier works which focus on the architecture-level design issues of NVM~\cite{lee:isca09, qureshi_micro09, ibm_isca09, pcm:isca2009}, such as wear-leveling, read-write disparity issues, etc. Some of those studies~\cite{cluster17:algorithm_nvm,hpdc16:wu,eurosys16:dulloor,unimem:sc17} use NVM as a replacement of DRAM at the architecture level. Many studies also consider NVM as a storage device, such as Onyx~\cite{onyx:hotstorage'2011}, Moneta~\cite{moneta:micro'2010}, and PMBD~\cite{msst14:chen}. Liu et al.~\cite{nas17:wei} investigate the current I\/O mechanisms on NVM but do not consider the effect of the file system. The Intel Optane product also uses NVM as a storage device. 

Some prior studies have explored file systems for NVM. For example, BPFS~\cite{sosp09:condit} uses shadow paging techniques for fast and reliable updates to critical file system metadata structures. SCMFS~\cite{sc11:wu} adopts a scheme similar to page tables in memory management for file management in NVM. PMFS~\cite{intel_pmfs} allows using memory mapping (mmap) for directly accessing NVM space and avoids redundant data copies. Some programing models for NVM are developed to take the benefit of byte-addressability and persistence. For example, Mnemosyne~\cite{mnemosyne_asplos11} gives a simple programming interface for NVM, such as the interface to declare non-volatile data objects. CDDS~\cite{Venkataraman:fast11} attempts to provide consistent and durable data structures. NV-Heaps~\cite{nv-heaps_asplos11} gives a simple model with support of transactional semantics. SoftPM~\cite{guerra:usenix2012} offers a memory abstraction similar to~\verb+malloc+ for allocating objects in NVM. 

In this study, we use Optane as a storage device and deploy both conventional file system and parallel file system atop for HPC applications. According to our experimental results, we find that the high-speed Intel Optane could significantly improve HPC application performance on both regular NFS and PVFS. However, the end-to-end effect is workload dependent and related to a variety of factors in the  I/O stack. 

\subsection{PVFS}
There are also some prior works studies the PVFS performance. For example,~\cite{pvfs2:PDP07} analyze the performance of PVFS2 under different hardware deployment and explores the performance bottleneck;~\cite{pvfs2gapfs:icpp2008} tries to leverage additional local storage to boost the local data read performance for parallel file systems;~\cite{mpiiopvfs:euro-par01} improves performances of collective I/O of MPI-IO on PVFS by avoiding the data-redistribution.

%% file: text/Conclusion_and_future_work.tex
\section{Conclusions}
We study the performance impact of the recent Intel Optane with HPC I/O intensive applications. Given the distinct performance characteristics of Optane, some of existing I/O stacks must be re-examined to optimize performance. According to our experimental results, we find Optane device has very high bandwidth but needs improve scalability under certain conditions; Collective I/O optimization does not always work. In the future, we may build a model to choose appropriate I/O method based on application workload characteristics and storage performance. Our work lays some foundation for the deployment of NVM  products in the future HPC.